\documentclass[fleqn,usenatbib]{mnras}

\usepackage{newtxtext,newtxmath}

\usepackage[T1]{fontenc}

\DeclareRobustCommand{\VAN}[3]{#2}
\let\VANthebibliography\thebibliography
\def\thebibliography{\DeclareRobustCommand{\VAN}[3]{##3}\VANthebibliography}

\usepackage{graphicx}	
\usepackage{amsmath}	

\title[Cross-Matching for Microlensing Stars]{Cross-Matching of OGLE, GAIA, and Hubble Catalogs: Evaluating the Probability of Resolving Lens Stars in Microlensing Events}

\author[S. Mozaheb, S. Rahvar]{
	S. Mozaheb,$^{1}$\thanks{E-mail:saeed.mozaheb@physics.sharif.edu}
	S. Rahvar,$^{1 , 2}$\thanks{E-mail:Rahvar@sharif.edu}
	\\
	$^{1}$Department of Physics, Sharif University of Technology, Azadi Ave, Tehran 11365-9161-Iran\\
	$^{2}$Research Center for High Energy Physics, Sharif University of Technology, Tehran, Iran
}
\date{Accepted XXX. Received YYY; in original form ZZZ}

\pubyear{2023}

\begin{document}
	\label{firstpage}
	\pagerange{\pageref{firstpage}--\pageref{lastpage}}
	\maketitle
	
	\begin{abstract}
	This study commenced by cross-matching data from the GAIA and OGLE telescopes with the aim of resolving the source star, long after microlensing is finished. The aim is breaking degeneracy between parameters of the microlensing equation, and ultimately calculating the mass of lens. We have examined different catalogs and found no evidence. \\
	Subsequently, employing the Monte Carlo method and guided by sensible assumptions, we embarked a simulation to discern the distribution of angular separation and to probe the feasibility of detecting this phenomenon. The results revealed that a mere $0.029\%$ of gravitational microlensing events exhibited separations exceeding $50$ milliarcseconds. Consequently, the likelihood of observing this phenomenon utilizing the OGLE and GAIA telescopes appears exceedingly far available. However, it is worth noting that instruments with very high angular resolution in the range of several tens of milliarcseconds present a viable avenue for such observations. Finally, we proposed $60$ microlensing events for which the observation of separation is more probable based on the measured proper velocity.
	\end{abstract}
	
	
	
	
	\section{INTRODUCTION}
	
	General relativity introduced by Albert Einstein Where gravitational lensing was one of predictions of this theory \citep{einstein1936lens}. This phenomena, was observed by Eddington and strongly supported the theory's validity \citep{eddington1919total}. Gravitational lensing is categorized into three types: strong, weak, and microlensing. Microlensing events are characterized by the inability to resolve the lensed image. These events are identified through the analysis of their light curves and the observable effect of amplification. \citep{schneider2006gravitational,paczynski1986gravitational}. This work focuses on observing the separation between the lens and source stars in gravitational microlensing.\\
	Calculating the mass of various objects is a challenging problem in astrophysics. Natural phenomena like gravitational lensing are crucial for mass calculation specially for dark objects and also, investigating the distribution of dark matter in galaxies \citep{paczynski1986gravitational}. While the notion of MACHOs serving as potential dark matter candidates has been ruled out, gravitational microlensing has subsequently been employed as a versatile astrophysical technique to investigate distant stars \citep{rahvar2015gravitational,paczynski1986gravitational} and detect exoplanets orbiting the lensed stars \citep{gaudi2010exoplanetary}. \citep{wyrzykowski2016black} identified $13$  microlensing events which are consistent with having dark lens. \\
	Determining mass solely through the light curve of microlensing events is not possible due to degeneracy in terms of Einstein crossing time. Therefore, various observational disturbances such as the parallax effect \citep{gould1994macho,rahvar2003study}, finite size effect \citep{mroz2020free}, frequency shift effect \citep{PhysRevD.101.024015} and source separation are utilized to break degeneracy and calculate mass \citep{gould2014microlens,rahvar2015gravitational}. This idea needs high precision telescopes to resolve the lens and source after a long time ending microlensing event. The relative proper motion may provide possibility to resolve them as two objects. \\
	Initially, due to recent research on simulations that calculate probability of observation of source-lens separation \citep{dehghani2022cross}, this work aims to observe the separation of microlensing events by cross-match data from the GAIA and OGLE telescopes. Also, Hubble Space Telescope data is used for higher resolution observations. Despite creating a comprehensive database of microlensing events, the source-lens separation was not observed in any of the data.
	To determine the probability of observing separation, simulations using the Monte Carlo method is performed. The probability of observing separation was investigated for various telescopes and projects, including Hubble and ALMA. The final conclusion indicates that a resolution of at least $50$ mas is necessary to separate the source and the lens. Ground telescopes that use adaptive optics can be very effective for investigating this separation. This observation was made once by \citep{bennett2020keck} using Keck telescope which helped calculating the mass of the central star in a planetary system. Required resolution exceeds the capabilities of the GAIA and OGLE telescopes, making it practically impossible to observe the lens and the source in the images captured by these telescopes.\\
	Considering the different conditions required for separation visibility, the number of visible events for different observational instruments is examined. As a result, the Hubble telescope and the ALMA radio telescope were identified as the best choices for future separation observations.\\
	In Section (\ref{sec:Deg Prob.}), we address the degeneracy problem and the calculation of lens mass. Additionally, we explore the consideration of perturbations, which aids in breaking the degeneracy. Moving to Section (\ref{sec:OGLE, GAIA}), we present the cross-match method that we implemented to identify the source-lens separation. Section (\ref{Simulation}) delves into the simulations conducted to estimate the probability of resolving the desired separation, taking into account the observational capabilities of available telescopes. conclusion is given in (\ref{sec:conclusion}).
	\section{GRAVITATIONAL MICROLENSING AND DEGENERACY PROBLEM} \label{sec:Deg Prob.}
	The amount of light deviation causes by a point mass is $ \alpha = 4 G M / bc^2 $ which can be calculated by solving geodesic in Schwarzschild metric. Here, the mass of the lens is $M$, $c$ is speed of light and the light impact parameter is $b$.

	\begin{figure}

	\includegraphics[width=\columnwidth]{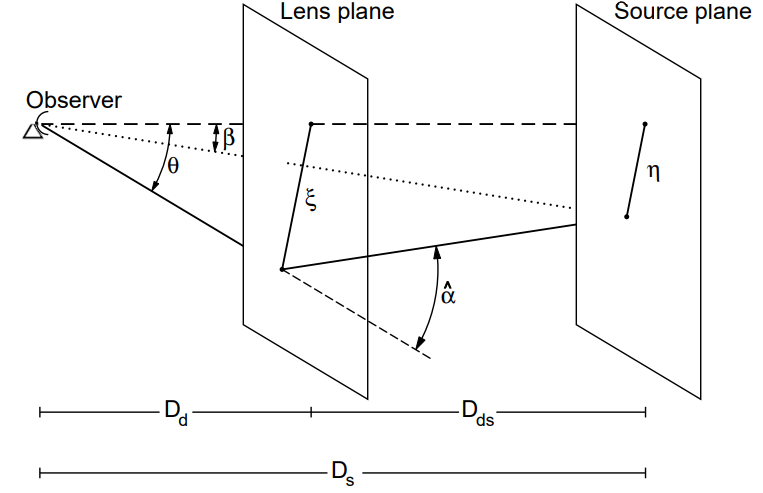}
		\caption{Illustration of the parameters involved in a gravitational lensing event. Here, $D_d$ represents the distance between the lens and the observer, while $D_s$ represents the distance between the source and the observer. \citep{bartelmann2001weak}}
		\label{fig:1}
	\end{figure}
	For gravitational lensing events, the Einstein angle is defined as the characteristic angle of the lens, which is determined by the mass of the lens $M$, the source distance $D_s$, and the ratio of the lens distance to the source distance $x$, these parameters are represented in Figure (\ref{fig:1}).  The Einstein angle is represented by the parameter $\theta_E$:
	\begin{equation}
		\theta_E = 0.901(\frac{M}{M_{\odot}})^{1/2}(\frac{D_s}{10kpc})^{-1/2}(\frac{1-x}{x})^{1/2} mas
	\end{equation}
	The Earth's movement, along with the source and the lens, causes the lensed light to appear to move from the perspective of the observer on Earth. This relative tangential velocity parameter is significant in determining Einstein crossing time. The relative angular velocity is:
	 \begin{equation}
	\mu = \frac{V_{S\perp} - V_{E \perp}}{D_s} - \frac{V_{L\perp} - V_{E\perp}}{D_d}
	\label{eq:2}
	\end{equation}
	Einstein crossing time is the specific time parameter of this event, defined as
	$t_E = \theta_E / \mu$.
	So finally, we can write Einstein crossing time:
	\begin{equation}
		\begin{split}	
			t_E & = 45.6 day (\frac{D_s}{8.5kpc})^{1/2}(\frac{M}{0.5 M_{\odot}} )^{1/2}(\frac{1-x}{x})^{1/2} \\
			& \times (|V_{S \perp} - V_{E \perp} - \frac{1}{x}(V_{ L \perp} - V_{E \perp})| \frac{1}{220km/s})^{-1} \label{eq3:29}\\
		\end{split}
	\end{equation}
	The relationship here clearly explains how mass is related to five other factors: By means that Einstein time $t_E$, as the observable parameter depends on source distance $D_s$, distance ratio $x$, source tangential velocity $ V_{S \perp} $ , and lens tangential velocity $ V_{L \perp}$. In gravitational microlensing events, light curve diagrams can only be used to calculate Einstein's time out of these parameters. To accurately determine the mass of the lens, it is essential to account for the interdependence among the known parameters referred to as degeneracy. Several methods can be employed to tackle this issue effectively. In specific instances of gravitational microlensing events, discrete disturbances such as parallax effect \citep{waagaard2017parallax}, Xallarap \citep{rahvar2008detecting}, and finite-source effect manifest \citep{choi2012characterizing,golchin2020measuring}. By meticulously observing the effects of these perturbations, an additional equation can be derived that establishes a connection between the pertinent parameters, thereby facilitating a more straightforward analytical solution. Consequently, through the meticulous assessment of the fluctuations in the said parameters, a reliable estimation of the lens mass can be achieved \citep{alcock1995first,waagaard2017parallax}.
	\subsection{Lens-Source Separation Detection}
	The methodology employed in this research to address the degeneracy problem entailed the observation of the separation between the lens and the source. By examining the location of this event several years after the occurrence of the gravitational microlensing event and being able to detect the source and the lens as distinct entities, it becomes feasible to compute the astrometric and astrophysical parameters for each of them individually. This approach facilitates a more precise estimation of the parameters associated with degeneracy and enables a more accurate calculation of the mass involved as investigated in \citep{bennett2008detection} and implemented in \citep{bennett2020keck}.
	
	In previous computer simulation studies \citep{dehghani2022cross}, extensive investigations have been conducted to explore the feasibility of observing the separation phenomenon. The research anticipated that a substantial number of separations could be detected by cross-match the microlensing observed through the OGLE ground telescope with the GAIA space telescope. Building upon this premise, the mass calculation was performed.
	
	Based on these findings, the present study commenced by cross-match two catalogs. Subsequently, a meticulous search was conducted within these comparison to identify the desire separation, which will be elaborated upon in the following section in detail

	\section{OGLE, GAIA and Hubble Cross-match} \label{sec:OGLE, GAIA}
	
\begin{figure}
		\includegraphics[width=\columnwidth]{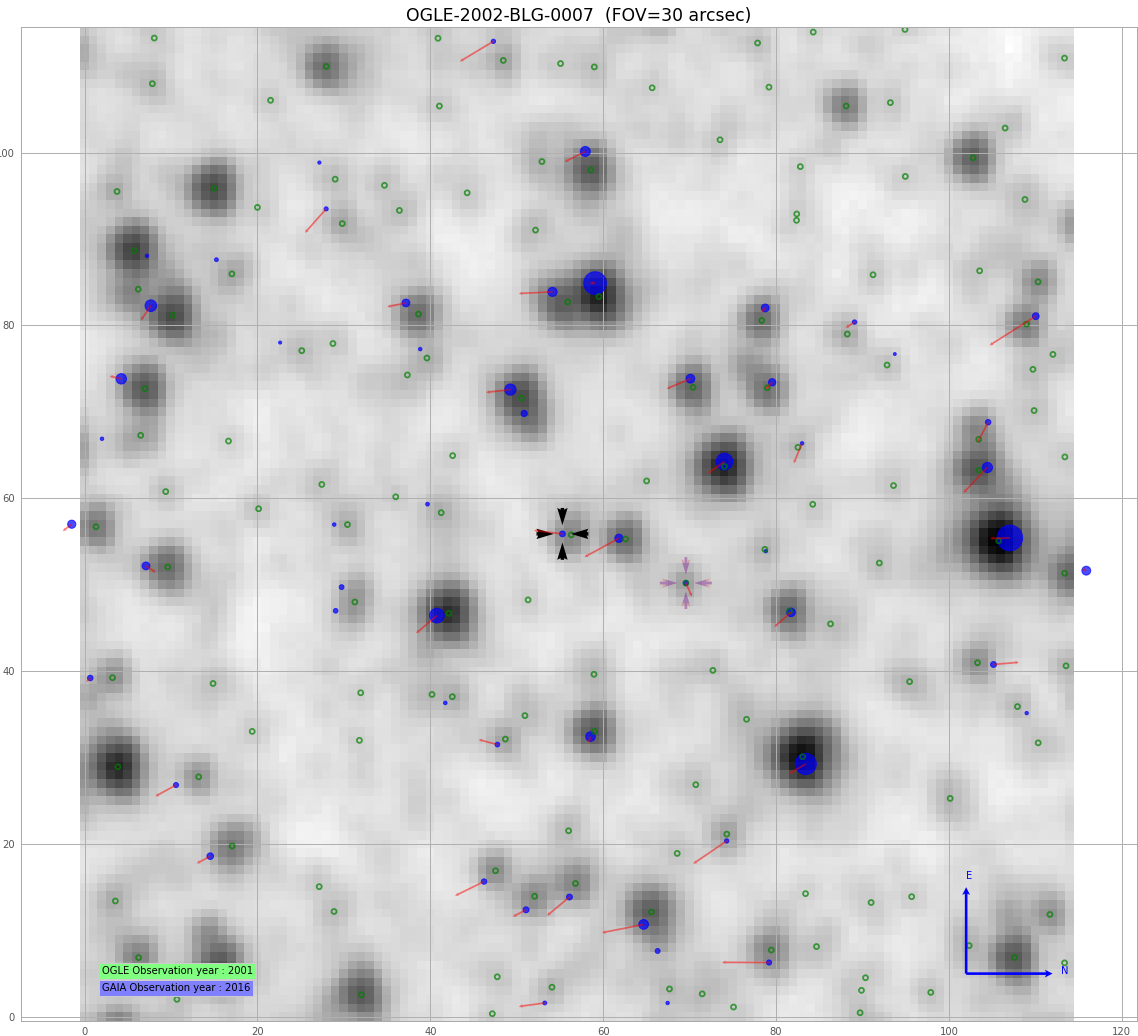}
		\caption{An example of cross-matched images: The background image is an OGLE image with its sources marked by green circles. The foreground image consists of GAIA-generated data, where each source is depicted with a velocity vector for illustration tangential proper motion, scaled up by a factor of $10$. The source which indicated with black arrows represents gravitational microlensing event. The source marked with faded red arrows is used for coordinate calibration.}
		\label{fig:2}
	\end{figure}

	In this research, the data from OGLE and GAIA were utilized. By cross-matching the light sources identified by the OGLE telescope as gravitational microlenses with the light sources in the GAIA catalog, access to all the recorded data from both projects was obtained. Figure (\ref{fig:2}) shows an example of the created cross-match.
	
\begin{figure}
		\centering
		\includegraphics[width=0.5\columnwidth]{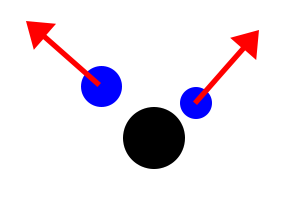}
		\caption{What is expected to be observed in image matching is for it to be considered as separation of lens and lens.}
		\label{fig:3}
	\end{figure}

	The search to determine the separation was conducted by analyzing each individual image. For the OGLE-related images, Gaussian point-spread function was applied using the DAOstarfinder programming function to identify the light sources and determine their exact locations. Additionally, vectors representing the specific velocities in the direction of Right ascension and Declination were plotted for each light source in the GAIA images. As the OGLE telescope images were not astrometrically calibrated and only prepared for photometric measurements, the star with the least specific motion was chosen as a guide for alignment. This star is indicated in red in the matching images. Subsequently, each matching image was examined to identify a pattern similar to Figure (\ref{fig:3}). The OGLE project recorded a total of about $21000$ gravitational microlensing events between $2002$ and $2022$. In the matching done by us, $14000$ events were matched successfully.\\
	The OGLE images were expected to show a single light source due to the separation, but the GAIA images captured several years later revealed two light sources moving away from each other. These two observed light sources can be identified as candidates for the source and the lens. Taking into account that the GAIA telescope had calculated the velocities of both sources, it became possible to calculate the mass.
	
	This process was repeated for every event in the OGLE project. However, in some events where the observation magnitude was over $20$, the matching was not performed accurately. Ultimately, the collation achieved a $91\%$ accuracy rate, meaning that $91$ out of $100$ events were correctly matched. Nonetheless, no separation was observed in the investigation of these events. Consequently, the data from OGLE is cross-matched with the Hubble data.
	
	After this analyzing GAIA data, we turned to Hubble Space Telescope data in an attempt to observe the separation between lens and sourve. Given its exceptional resolution and the availability of its raw data to the public, the Hubble Space Telescope was a compelling choice for this endeavor. However, due to its limited field of view, mainly targeting non-stellar light sources, the telescope observed very few gravitational microlensing events initially recorded by OGLE in later years.
	
\begin{figure}
	\includegraphics[width=\columnwidth]{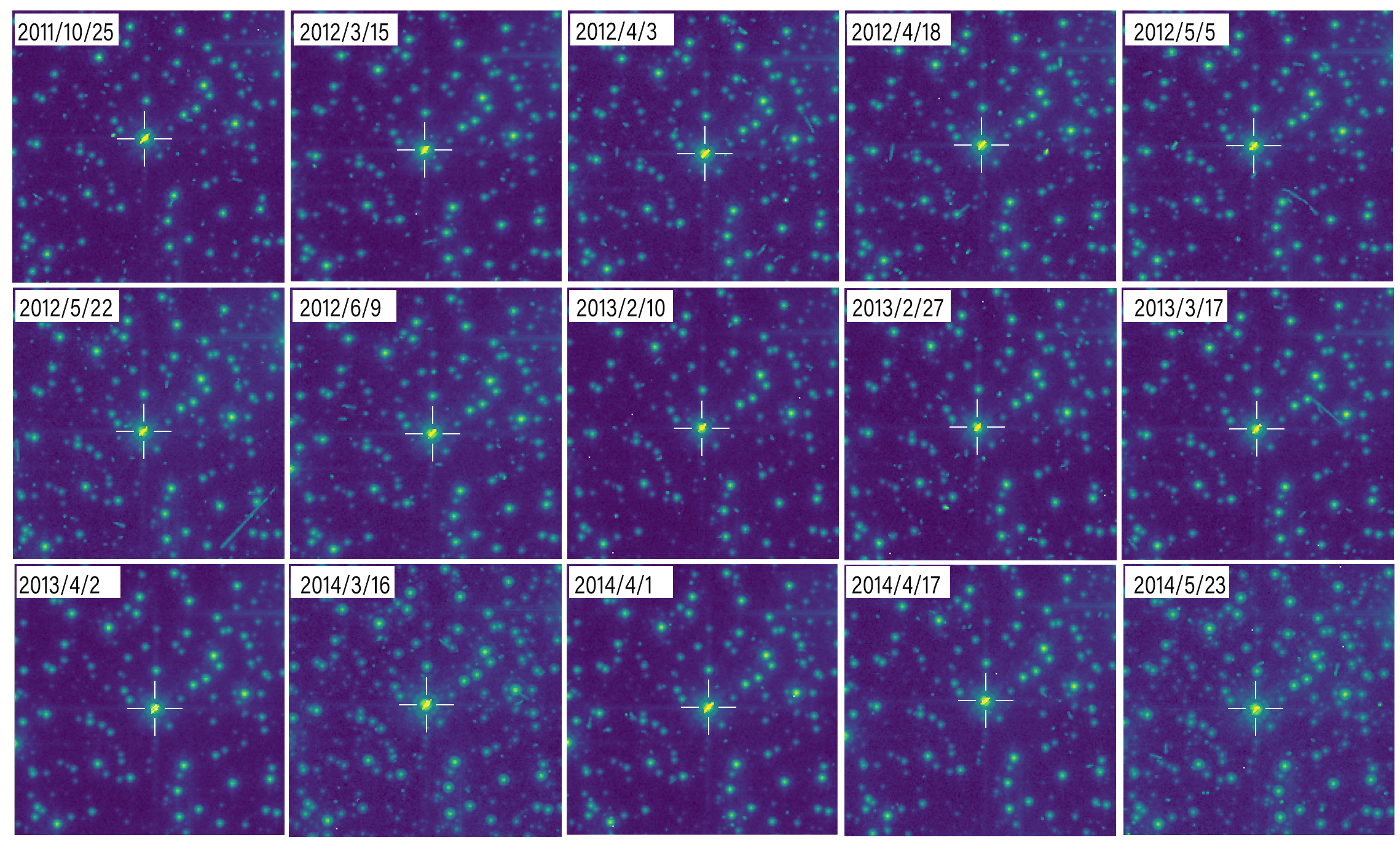}
	\caption{
		This is a time series of images taken by the Hubble Space Telescope of $OGLE-2014-BLG-0265$. The event reached its maximum brightness on March $19$, $2014$, with $12$ images captured before the event and $3$ taken after. However, despite our efforts, we did not identify any significant movement indicative of the desired separation and there is no other images from this field of view in Hubble archive.  }
	\label{fig:4}
	\end{figure}

	Observing gravitational microlensing in a single Hubble image does not allow for direct mass calculation. Instead, it necessitates multiple images captured in different years to analyze the movement of objects within them and estimate relative velocities. Consequently, we were able to identify only two gravitational microlensing events ($OGLE-2014-BLG-0265$ and $OGLE-2011-BLG-0123$)  that Hubble had imaged over different years. A time series of images related to one of these events is presented in Figure (\ref{fig:4}).
	After examining the Hubble data, we doubted the results of the previous simulations and decided to do this simulation again to get more accurate results.
	
	\section{Lens-Source Simulation} \label{Simulation}
	To assess the likelihood of separation visibility in gravitational microlensing events, given the multitude of kinematic and astrophysical factors influencing this measurement, we employ the Monte Carlo method to calculate this probability. This method necessitates deriving the variances of the various parameters involved and subsequently generating random gravitational microlensing events based on these parameters.
	
	In our simulation, we assume that gravitational microlensing is initiated by one source and one lens star. It's important to note that other celestial objects, such as black holes, can also play a role in the occurrence of these events. We further assume the source star's location in the central bulge of the Milky Way galaxy, while the lens star is positioned in the galactic disk.
	
	For sampling, we select an area with a 4-degree arc diameter around the bulge at the galaxy's center, based on the distribution of observed gravitational microlensing event coordinates from the OGLE telescope.
	\begin{figure}
		\includegraphics[width=\columnwidth]{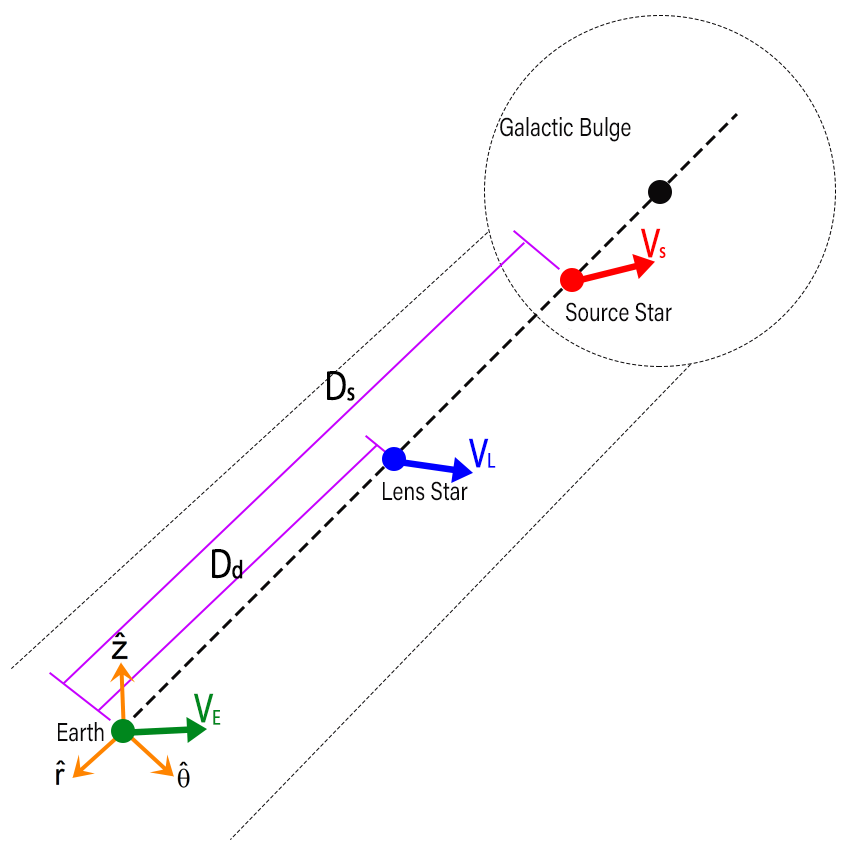}
		\caption{A schematic diagram depicting the positions of the three primary members of this event, along with an explanation of the cylindrical coordinate system employed.}
		\label{fig:5}
	\end{figure}
	Subsequently, using GAIA data, we collect information on the distances between the stars within this region. The spatial distribution of stars in this field of view obtained from \citep{moniez2017understanding}. Following this, we apply the kinematic parameters, the definitions of which are elucidated in Figure (\ref{fig:5}). We formulate both a global tangential and a dispersion set of sentences for these parameters, encompassing the sun, the source star, and the lens star: 
	$ \vec{V} = \vec{V}_{global} + \vec{V}_{dispersion} $.\\
	For the Sun, the global term pertains to $220 km/s$ , accompanied by the dispersion term, representing the Sun's motion at the speed of $16.5 km/s$ towards coordinates defined by $l_0=25^\circ$ and $b_0=53^\circ$.\\
	Concerning the lens star located within the galactic disc, the general term relates to the star's distance from the galaxy's center, measured at $V_{L,\theta} = V_{R_0,\theta}(\frac{r}{R_0})^{\beta}$ with $V_{R_0,\theta} = 220 km/s$ and $R_0 = 8 kpc$ distance of sun from center of galaxy. The dispersion velocities in various directions for lens star located in galactic disk are as follows:
	\begin{equation}
		\begin{split}
		& \bar{\sigma}_r = 39.4 \pm 0.3 km/s  \\
		& \bar{\sigma}_{\theta} = 25.21 \pm 0.3 km/s \\
		& \bar{\sigma}_{z} = 21.1 \pm 0.2 km/s \\
		\end{split}
	\end{equation}
	For the source star situated in the galactic bulge, the general velocity is $V_{S,\theta} = -19km/s $, and the dispersion velocity in all directions remains consistent at:
	\begin{equation}
			\bar{\sigma}_{r,\theta,z} = 104 \pm 10  km/s \\
	\end{equation}
	Also, the time interval between the observations of two telescopes was considered based on the time interval of OGLE and GAIA observations with an average of $10$ years and a dispersion of $3$ years.
		
\begin{figure}
		\includegraphics[width=\columnwidth]{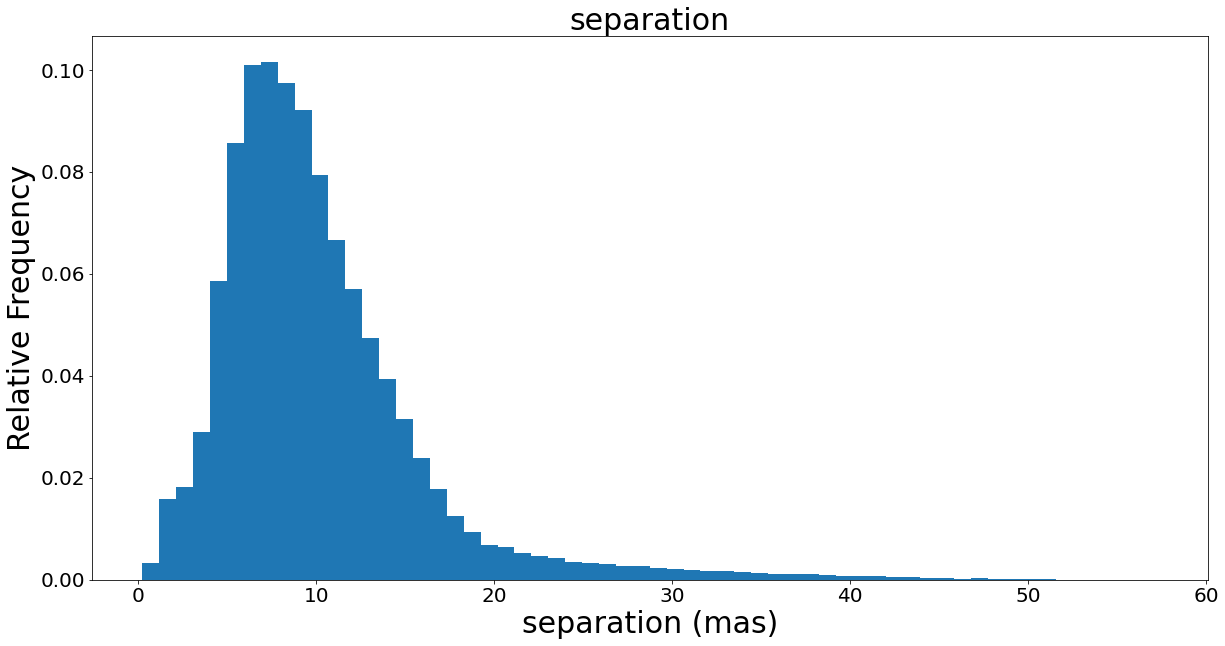}
		\caption{Normalized distribution of the separation parameter from 1 million simulation replications}
		\label{fig:6}
	\end{figure}

	This process led to the computation of equation (\ref{eq:2}) for each simulated event, which subsequently facilitated the determination of event separation rates. Refer to Figure (\ref{fig:6}) for an illustrative depiction of the separation results obtained through these simulations.
	To calculate the probability of observing the separation, various parameters come into play, including the likelihood of detecting a gravitational microlensing event and the received brightness from this event.
	
\begin{figure}
	\includegraphics[width=\columnwidth]{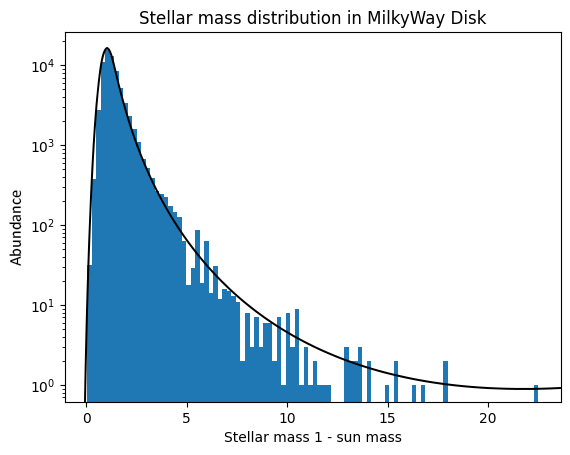}
	\caption{Stellar mass distribution in galaxy disk based on Besançon Model of the Galaxy\citep{10.1007/978-3-642-18418-5_18}} 
	\label{fig:7}
	\end{figure}

	As outlined by \citep{alcock2001macho}, microlensing detection efficiency is related to Einstein crossing time of the event (i.e $E=E(t_E)$). To incorporate this into our simulation, it becomes imperative to assign a corresponding value related to the mass of the lens stars. We accomplished this by utilizing Bosanson simulation, and the outcome for disc stars is depicted in Figure (\ref{fig:7}).
	
\begin{figure}
	\includegraphics[width=\columnwidth]{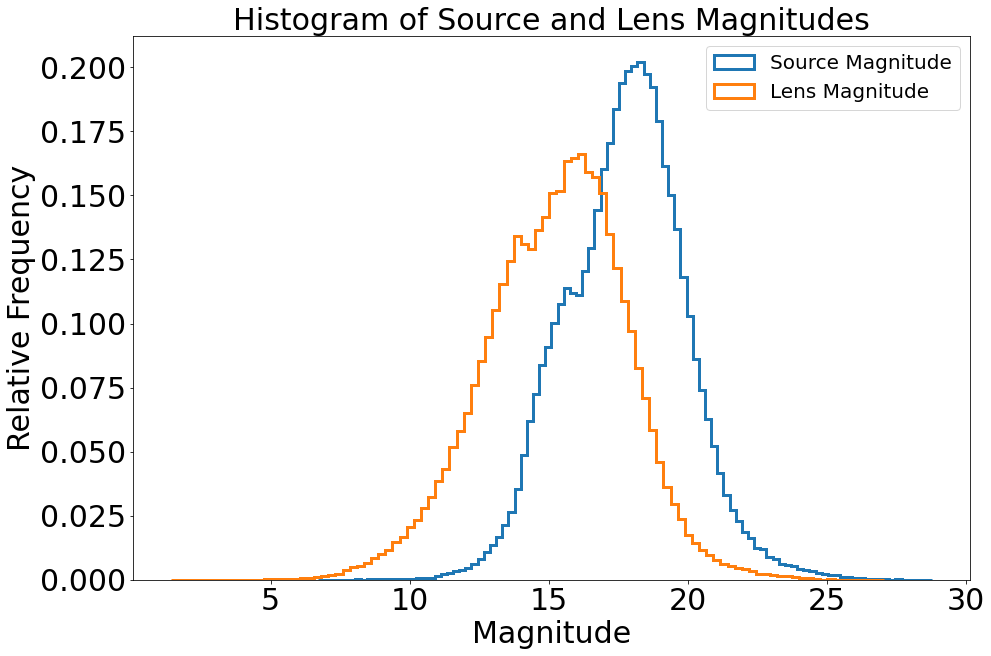}
	\caption{Source and Lens brightness magnitude based on absolute magnitude of stars and distribution of stars distances from observer}
	\label{fig:8}
	\end{figure}

	Another important factor in determining the visibility of a gravitational microlensing event is the apparent brightness of star. Drawing upon the luminosity of the light sources found in the GAIA catalog and considering their distances calculated from parallax data, we were able to construct absolute magnitude for stars in both the galactic disc and bulge. Consequently, we allocated a brightness proportional to each of the light sources. The brightness distributions for the source and lens stars are represented in Figure (\ref{fig:8}).
	
	 \section{Conclusion}\label{sec:conclusion}
\begin{figure}
		\includegraphics[width=\columnwidth]{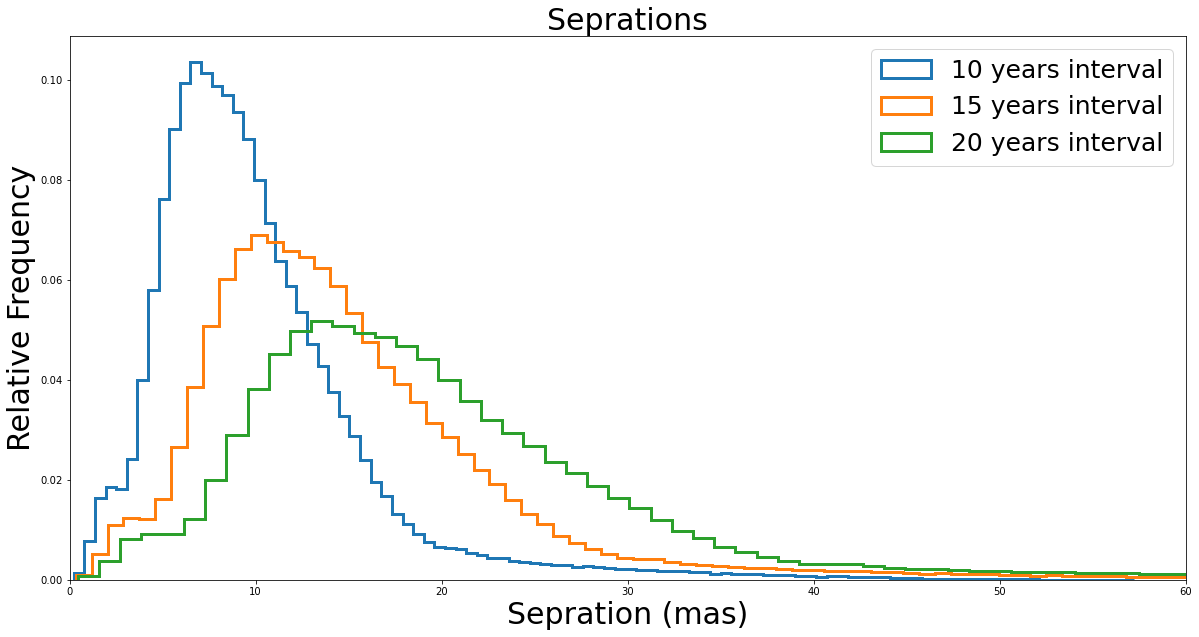}
		\caption{Comparison of separation parameter for 10, 15 and 20 years time intervals} 
		\label{fig:9}
	\end{figure}

	Based on the simulation results (figure(\ref{fig:6})), it is evident that a mere $0.029\%$ of gravitational microlensing events produce a separation exceeding $50$ milli arc seconds. Out of this fraction, only $24\%$ exhibit a magnitude below $20$ and can be observed with a reasonable probability from Earth. This calculation is based on the distribution of event detection interval by OGLE and data recording of GAIA(average of 10 years with dispersion of 3 years). Meanwhile, after $5$ years this proportion increase considerably. Events with over $50$ mac separation will account for $1.18\%$ and over $30.23\%$ of such events would be visible due to brightness limitation. These values will be $3.16\%$ and $30.67\%$ for the next $5$ years ($20$ years after microlensing detection). The corresponding comparison is shown in figure (\ref{fig:9}) . This implies that in the coming years, cross-matching the latest observational data with the OGLE catalog could make it more likely to detect separation between source and lens. However, with the current data, detection of this phenomenon is very unlikely. The threshold of $50$ milliseconds of arc, which is just slightly below the highest resolution of the Hubble Space Telescope, underscores the considerable challenge in detecting the separation phenomenon, more so than previously anticipated. This leads to the primary conclusion that the odds of observing this separation, by combining OGLE and GAIA datasets, appear exceedingly unlikely. This conclusion is corroborated by the absence of separations found in the data matching process.
	Similar results were obtained in other research \citep{bennett2020keck}. This study utilized the resolution of a source-lens event to calculate the mass of the lens, which, in this case, was a star with a Neptunian planet. The researchers derived the mass ratio from light curves, allowing them to estimate the mass of both the star and the planet. In their investigation, an image was captured using the Keck telescope $14$ years after the microlensing event, revealing a $55$ milliarcsecond separation between the lens and the star in the final images.
	
	 \citep{delplancke2001resolving} explored the possibility of observing gravitational microlensing with VLTI, it becomes evident that the current prospects for observing the separation phenomenon are predominantly confined to data with exceptionally high resolutions, such as VLTI and ALMA. It is worth noting that the observation of separations in radio wavelengths is restricted to cases where both the source and lens exhibit substantial radio radiation. While stars like Mira and T-Tauri within the galaxy emit notable radio radiation, their population remains relatively limited. Previously \citep{karami2016resolving} has studied gravitational microlensing for these objects. 
	
	Considering a constant velocity for stars in the Milky way within our observation time-scale, it is anticipated that observations with extended time intervals (in the range of $15$ to $20$ years between two observations) in the upcoming years will significantly elevate the likelihood of observing the separation phenomenon.\\
	Ultimately, we recommend $60$ events identified as potential candidates for resolving in gravitational microlensing events in tables \ref{tab:1} and \ref{tab:2}. These selections stem from cross-matching data between GAIA and OGLE, with proper motion serving as the decisive parameter. we choose $15$ events from OGLEIII and $45$ from OGLEIV.

	 \section*{DATA AVAILABILITY}\label{sec:data-Availability}
	 
	The data underlying this article and cross-match results are available in Microlensing Separation Detection Repository, at \href{https://github.com/saeedmozaheb}{https://github.com/saeedmozaheb}
	
	
	
	\bibliographystyle{mnras}
	\bibliography{example.bib} 

	
	
	
	\appendix

	\section{Introducing Candidates}
	
	In this section, we introduce gravitational microlensing events in which separation is more likely to be observed. Based on the cross-match between GAIA and OGLE catalogs, the proper motion of the events were determined and the events with proper velocities greater than $ mean + 3\sigma $ were introduced. 
			
	\begin{table*}
		\caption{Top candidates of resolving in OGLEIII based on proper motion detected by GAIA space telescope.}
		\label{tab:1}
		\begin{tabular}{|l|l|l|l|l|l|l|}
			\hline
			OGLE Name & GAIA ID  & $Ra_{deg}$ & $Dec_{deg}$ & $\mu_{(mas/yr)}$ & G Mag $^a$ & $t_{E(day)}$ \\
			\hline
			2004-BLG-161 & $4062741369288060800$ & $269.62878$ & $-28.180710$ & $36.48 \pm 2.30$ & $20.10$ &  $ 5.89 \pm 0.47$  \\
			2002-BLG-063 & $4067963985105597824$ & $265.62533$ & $-25.159622$ & $24.35 \pm 2.61$ & $20.38$ &  $ 13.56 \pm 1.29$  \\		
			2006-BLG-224 & $4062826860616038784$ & $271.58647$ & $-28.195817$ & $22.87 \pm 1.21$ & $19.36$ & $1.49 \pm 0.03$ \\
			2009-BLG-091 & $4056521535135350400$ & $269.14828$ & $-29.388038$ & $20.18 \pm 1.34$ & $18.94$ & $39.00 \pm 0.64$ \\
			2009-BLG-113 & $4044141205918658176$ & $270.40029$ & $-30.722650$ & $20.01 \pm 0.46$ & $15.61$ & $14.32 \pm 0.46$ \\
			2003-BLG-350 & $4068114961855612032$ & $266.09246$ & $-24.571951$ & $19.20 \pm 1.08$ & $19.72$ & $14.00 \pm 0.44$ \\
			2008-BLG-242 & $4062651450009511424$ & $269.47422$ & $-28.131145$ & $18.87 \pm 1.03$ & $19.21$ & $13.44 \pm 0.07$ \\
			2002-BLG-350 & $4041646208000844160$ & $267.68555$ & $-34.087184$ & $17.71 \pm 1.25$ & $19.90$ & $24.68 \pm 3.64$ \\
			2002-BLG-337 & $4042496645812141184$ & $269.33007$ & $-33.511307$ & $17.63 \pm 0.59$ & $18.80$ & $12.25 \pm 0.23$ \\
			2007-BLG-590 & $4043225828379478016$ & $268.67603$ & $-33.162788$ & $17.30 \pm 1.75$ & $19.90$ & $10.34 \pm 0.50$ \\
			2005-BLG-305 & $4056287579617814912$ & $268.67116$ & $-30.072743$ & $17.16 \pm 0.77$ & $18.32$ & $13.39 \pm 2.84$ \\
			2008-BLG-408 & $4062484118116292608$ & $270.16814$ & $-28.506867$ & $17.08 \pm 1.43$ & $19.14$ & $70.21 \pm 11.41$ \\
			2008-BLG-613 & $4062615788792334464$ & $269.13666$ & $-28.519895$ & $17.01 \pm 0.66$ & $18.75$ & $45.45 \pm 1.18$ \\
			2007-BLG-552 & $4050406983474790400$ & $271.58890$ & $-28.839279$ & $16.94 \pm 1.06$ & $19.00$ & $11.37 \pm 0.23$ \\
			2002-BLG-272 & $4118048698256676736$ & $264.69169$ & $-21.111670$ & $16.92 \pm 1.45$ & $19.38$ & $2.53 \pm 0.18$ \\
			\hline
			\multicolumn{7}{l}{a: G-band: Gaia's white light magnitude (330 - 1050nm) \citep{brown2021gaia} }\\
			
		\end{tabular}
	\end{table*}
	\begin{table*}
		\caption{Top candidates of resolving in OGLEIV based on proper motion detected by GAIA space telescope.}
		\label{tab:2}
		\begin{tabular}{|l|l|l|l|l|l|l|}
			\hline
			OGLE Name & GAIA ID  & $Ra_{deg}$ & $Dec_{deg}$ & $\mu_{(mas/yr)}$ & G Mag $^a$ & $t_{E(day)}$ \\
			\hline
			
			2011-BLG-0061 & $4044073173675897472$ & $269.4517017$ & $-30.95517821$ & $112.89 \pm 0.17$ & $17.00$ & $882979.269$   \\
			2012-BLG-1027 & $5980774811978506624$ & $258.7892033$ & $-30.20803266$ & $37.09 \pm 1.06$ & $20.13$ & $2.93 \pm 0.082$ \\
			2013-BLG-0004 & $4068223121980873600$ & $265.4639788$ & $-24.45023805$ & $34.05 \pm 0.62$ & $18.99$ & $15.26 \pm 0.386$ \\
			2012-BLG-1213 & $4056294455913437696$ & $268.996156$ & $-29.87321965$ & $29.37 \pm 0.56$ & $19.14$ & $16.44 \pm 0.143$ \\
			2013-BLG-0005 & $4041751181184034944$ & $266.9550319$ & $-34.14068047$ & $29.30 \pm 1.64$ & $20.35$ & $27.54 \pm 0.718$ \\
			2012-BLG-0853 & $4057401178708838016$ & $267.7557356$ & $-28.70658233$ & $29.22 \pm 0.05$ & $15.97$ & $0.78 \pm 0.015$ \\
			2017-BLG-1044 & $4044201683435491200$ & $269.9074331$ & $-30.59145968$ & $26.69 \pm 0.36$ & $18.44$ & $0.88 \pm 0.029$ \\
			2014-BLG-1679 & $4062577851300138240$ & $269.6969118$ & $-28.65232268$ & $25.28 \pm 1.72$ & $19.43$ & $9.87 \pm 0.374$ \\
			2018-BLG-1730 & $4063532472163597568$ & $269.0610067$ & $-27.70846212$ & $24.24 \pm 0.88$ & $20.00$ & $4.33 \pm 0.274$ \\
			2016-BLG-0282 & $4041419055751492224$ & $267.168099$ & $-35.57943674$ & $24.20 \pm 0.96$ & $18.63$ & $18.86 \pm 0.074$ \\
			2013-BLG-0006 & $4062593935921180416$ & $269.7644697$ & $-28.4206839$ & $24.14 \pm 0.68$ & $16.67$ & $32.54 \pm 0.146$ \\
			2015-BLG-0620 & $4043398936841900544$ & $268.0827489$ & $-33.17631868$ & $23.57 \pm 1.53$ & $19.65$ & $24.639 \pm 0.84$ \\
			2011-BLG-0473 & $4062777619010136448$ & $270.1510563$ & $-27.77020718$ & $23.56 \pm 1.49$ & $19.50$ & $6.809 \pm 0.048$ \\
			2017-BLG-0381 & $4041474374998087040$ & $267.4616912$ & $-34.92498384$ & $23.35 \pm 2.11$ & $19.80$ & $72.284 \pm 3.16$ \\
			2015-BLG-1024 & $4043382199512780160$ & $269.4345185$ & $-32.26154717$ & $23.18 \pm 1.37$ & $19.79$ & $23.833 \pm 1.614$ \\
			2022-BLG-0060 & $4056327162008855680$ & $269.2863419$ & $-29.63845511$ & $21.87 \pm 0.08$ & $16.24$ & $16.371 \pm 0.56$ \\
			2017-BLG-0318 & $4043712052876957824$ & $270.4078286$ & $-32.17890951$ & $21.62 \pm 1.22$ & $19.80$ & $3.962 \pm 0.144$ \\
			2022-BLG-0260 & $4061103547960320512$ & $265.6300101$ & $-26.40105123$ & $20.91 \pm 2.16$ & $20.43$ & $44.253 \pm 1.486$ \\
			2017-BLG-1537 & $4041936857013164800$ & $266.8620549$ & $-33.46353572$ & $20.88 \pm 0.43$ & $19.05$ & $11.104 \pm 0.201$ \\
			2019-BLG-0239 & $4056491680794776704$ & $268.6880389$ & $-29.61538108$ & $20.35 \pm 1.06$ & $18.93$ & $24.769 \pm 0.225$ \\
			2014-BLG-1739 & $4043201673345585664$ & $268.1079637$ & $-33.45638128$ & $20.09 \pm 1.66$ & $19.67$ & $9.78 \pm 0.103$ \\
			2012-BLG-1393 & $4063554462406705536$ & $269.492422$ & $-27.45635038$ & $19.93 \pm 1.00$ & $18.30$ & $23.934 \pm 0.745$ \\
			2013-BLG-0008 & $4050954089398508672$ & $272.758165$ & $-27.86918009$ & $19.82 \pm 0.68$ & $18.59$ & $9.195 \pm 0.185$ \\
			2016-BLG-0542 & $4064909507432758272$ & $271.8802831$ & $-25.57526051$ & $19.75 \pm 0.97$ & $19.33$ & $12.965 \pm 0.668$ \\
			2016-BLG-1651 & $4060489161450409600$ & $263.9847059$ & $-28.19889371$ & $19.73 \pm 1.42$ & $19.83$ & $30.632 \pm 1.46$ \\
			2013-BLG-0010 & $4041939915021368576$ & $266.9596781$ & $-33.31566796$ & $19.55 \pm 0.99$ & $19.54$ & $20.842 \pm 0.762$ \\
			2011-BLG-0362 & $4062745531088556288$ & $269.5724639$ & $-28.0996355$ & $19.35 \pm 0.78$ & $17.87$ & $22.542 \pm 0.449$ \\
			2011-BLG-0945 & $4062675059292894464$ & $269.8576718$ & $-28.43470195$ & $19.28 \pm 0.20$ & $17.73$ & $8.58 \pm 0.115$ \\
			2017-BLG-1525 & $4065693802815477888$ & $271.8022592$ & $-24.99892613$ & $19.03 \pm 0.40$ & $18.89$ & $6.637 \pm 0.414$ \\
			2018-BLG-0810 & $4062612833786281088$ & $269.0584278$ & $-28.62470256$ & $18.89 \pm 0.49$ & $17.56$ & $0.661 \pm 0.015$ \\
			2018-BLG-1740 & $4055765139815537792$ & $267.945982$ & $-30.92866272$ & $18.84 \pm 1.38$ & $19.51$ & $21.816 \pm 0.403$ \\
			2014-BLG-2015 & $4041389682454535168$ & $265.8254412$ & $-34.72022203$ & $18.47 \pm 2.14$ & $20.24$ & $46.945 \pm 7.705$ \\
			2012-BLG-0054 & $4063305178372953088$ & $271.1995579$ & $-26.79606505$ & $18.38 \pm 1.10$ & $18.77$ & $17.928 \pm 0.086$ \\
			2019-BLG-1117 & $4056219100786930816$ & $269.0217936$ & $-30.30468202$ & $18.31 \pm 0.36$ & $18.30$ & $22.201 \pm 0.17$ \\
			2012-BLG-0790 & $4056616372275423360$ & $268.5480385$ & $-28.72121258$ & $18.28 \pm 0.60$ & $17.98$ & $8.752 \pm 0.168$ \\
			2018-BLG-0603 & $4061951855541710080$ & $265.0064516$ & $-25.56900354$ & $18.28 \pm 0.70$ & $19.40$ & $7.902 \pm 0.582$ \\
			2014-BLG-0095 & $4062332286515948928$ & $269.7500704$ & $-29.27495509$ & $18.26 \pm 1.00$ & $18.68$ & $11.246 \pm 0.214$ \\
			2016-BLG-0728 & $4061874443077112320$ & $265.4572736$ & $-25.94042816$ & $18.25 \pm 0.61$ & $18.38$ & $9.225 \pm 0.095$ \\
			2015-BLG-1362 & $4041546225426345088$ & $266.9678953$ & $-34.58893805$ & $18.05 \pm 1.44$ & $19.66$ & $121.285 \pm 22.763$ \\
			2013-BLG-0011 & $4117984960891575424$ & $264.2843569$ & $-21.85027407$ & $18.05 \pm 2.25$ & $20.21$ & $55.929 \pm 2.105$ \\
			2012-BLG-0390 & $4062800193314792832$ & $269.7567579$ & $-27.53970982$ & $17.86 \pm 0.41$ & $18.82$ & $19.24 \pm 0.518$ \\
			2015-BLG-0198 & $4067954330017995008$ & $265.4848581$ & $-25.34528433$ & $17.62 \pm 0.18$ & $17.74$ & $96.649 \pm 5.95$ \\
			2019-BLG-0672 & $4050420860487572352$ & $271.2847121$ & $-28.84668324$ & $17.60 \pm 1.17$ & $18.98$ & $12.122 \pm 0.112$ \\
			2011-BLG-0882 & $4041466128651375744$ & $267.6340073$ & $-35.02864029$ & $17.42 \pm 1.88$ & $19.79$ & $9.651 \pm 0.657$ \\
			2013-BLG-0013 & $4056250544149378944$ & $269.5584665$ & $-30.0362847$ & $17.41 \pm 1.86$ & $19.61$ & $3.156 \pm 0.063$ \\
			\hline
			\multicolumn{7}{l}{a: G-band: Gaia's white light magnitude (330 - 1050nm) \citep{brown2021gaia} }\\
		\end{tabular}
	\end{table*}


	\bsp	
	\label{lastpage}
\end{document}